\documentclass[12pt,a4paper,dvips]{article}
\usepackage{a4p}
\usepackage{cite,mcite}
\usepackage{graphicx}
\usepackage{epsfig}
\usepackage{physics }
\usepackage{l3_title,ifthen}
\journalname{Phys. Lett. B}
\date{March 29, 2000}
\preprint{2000-047}
\newlength{\capindent}
\newlength{\capwidth}
\newlength{\figwidth}
\newcommand{\icaption}[2][!*!,!]{\hspace*{\capindent}%
  \begin{minipage}{\capwidth}
    \ifthenelse{\equal{#1}{!*!,!}}%
      {\caption{#2}}%
      {\caption[#1]{#2}}
  \end{minipage}}
\newcommand{\LL}{\ensuremath{\mathrm{\ell^+ \ell^-}}}
\newcommand{\EE}{\ensuremath{\mathrm{e^+ e^-}}}
\newcommand{\EEG}{\ensuremath{\mathrm{e^+ e^- (\gamma)}}}

\newcommand{\MMG}{\ensuremath{\mathrm{\mu^+ \mu^- (\gamma)}}}
\newcommand{\ZZ}{\ensuremath{\mathrm{Z/\gamma^* Z/\gamma^*}}}
\newcommand{\QQ}{\ensuremath{\mathrm{q \bar{q}}}}

\newcommand{\QQG}{\ensuremath{\mathrm{q \bar{q} (\gamma)}}}

\newcommand{\zee}{\ensuremath{\mathrm{Z \EE}}}

\newcommand{\TTG}{\ensuremath{\mathrm{\tau^+\tau^- (\gamma)}}}
\newcommand{\FF}{\ensuremath{\mathrm{f\bar{f}}}}

\newcommand{\WW}{\ensuremath{\mathrm{W^+ W^-}}}

\begin{document}

\begin{titlepage}
\title{\boldmath Search for an invisibly decaying Higgs boson  in\\
$\epem$ collisions at \rts\ = 183 -- 189 \gev{}}

\author{L3 Collaboration}

\begin{abstract}
A search for a  Higgs boson decaying into invisible particles 
is performed using the data
collected at LEP by the L3 experiment at centre-of-mass energies of
$183\gev{}$ and $189\gev{}$.   
The integrated luminosities are respectively 55.3~\pb{} and 176.4~\pb{}.
The observed candidates are consistent with the expectations from Standard
Model processes.
In the hypothesis that the production cross section of this Higgs boson
equals the Standard Model one and the branching ratio into invisible
particles is 100\%, a lower mass
limit of $89.2\gev{}$ is set at $95\%$ confidence level.
\end{abstract}

\submitted
\end{titlepage}


\section{Introduction}
In some extensions of the Standard Model the Higgs boson can decay into stable weakly
interacting particles, thus yielding invisible final states~\cite{hinv_th}.
For example the minimal supersymmetric extension of the Standard
Model predicts that the Higgs boson can decay into a pair of invisible
neutralinos. 

A search is performed for a Higgs boson  produced through the Higgs-strahlung
process, $\EE\to \mathrm{Z^{*} \to hZ}$.
The Z boson decays into fermion pairs yielding two different investigated
topologies: 
two acoplanar jets plus missing
energy, corresponding to the Z boson hadronic decays, and two acoplanar charged leptons
plus missing energy, corresponding to 
decays of the Z boson into electrons or muons. 
Data collected by the L3 experiment \cite{l3det} at LEP
centre-of-mass energies of  
$\rts = 183 \gev{}$ and $189 \gev{}$ are analysed.
The corresponding integrated luminosities are respectively 55.3~\pb{}
and 176.4~\pb{}. Results at lower centre-of-mass energies 
have been reported by L3~\cite{Acciarri:1997hinv} and 
by the other LEP experiments~\cite{ALEPH:183}.


\section{Event simulation}
To determine the signal efficiency,  samples of Higgs boson events are
generated using the PYTHIA Monte Carlo program \cite{pythia} for masses
between $55\gev{}$ and $100\gev{}$.

For the background studies the following Monte Carlo programs are used:
PYTHIA ($\EE \to \QQG,\ \EE \to \ZZ$ and $\EE \to \zee$),
KORALW \cite{koralw} ($\EE \to \WW$),
KORALZ \cite{koralz} ($\EE \to \MMG$, $\EE \to \TTG$), 
PHOJET \cite{phojet} ($\EE \to \EE \QQ$),
DIAG36 \cite{diag36} ($\EE \to \EE \LL$),
BHWIDE \cite{bhwide} ($\EE \to \EEG$),
and EXCALIBUR \cite{excalibur} for the other four-fermion final states.
For each centre-of-mass energy, the number of simulated
background events 
corresponds to at least 50 times the
number of expected events except for
the two-photon interactions ($\EE \to \EE \FF$)
and Bhabha scattering ($\EE \to \EE$) for which twice and seven
times the collected luminosity are simulated, respectively.

The L3 detector response is simulated using the GEANT~3.15
program \cite{geant}, which takes into account the effects of energy loss,
multiple scattering and showering in the detector. The GHEISHA program
\cite{gheisha} is used to simulate hadronic interactions in the detector.
Small time-dependent inefficiencies of the different subdetectors are  taken
into account in the simulation procedure.


\section{Search in the hadronic channel}

A cut-based analysis is used to select events
in the hadronic channel. After a common preselection
two sequential  selections are separately optimised
for light (below $80 \gev{}$) and heavy (above
$80 \gev{}$) Higgs boson masses.
Unless otherwise stated,  the events
are constrained to have two jets using  the DURHAM algorithm~\cite{DURHAM}.

\subsection{Preselection}

High-multiplicity hadronic events are selected at $\sqrt{s} =
189\gev{}$. Events coming
from QCD processes and hadronic decays of W and Z boson pairs are rejected
by requiring  a missing momentum larger than $10 \gev{}$. The absolute values
of the cosine
of the polar angle of the jets and of the missing 
momentum vector have to be less than 0.9, to reject
events with a high-energy initial-state radiation photon emitted close
to the beam axis.
In addition, events with large energy depositions in the forward calorimeters are vetoed to
reduce the background contribution from  the $\zee$ and $\QQG$
processes  and from residual two-photon interactions. Coplanar events are
rejected to further suppress these last processes.

Events with energetic and isolated charged leptons are removed to
decrease the contamination from semi-leptonic decays of W boson
pairs. This cut is designed to keep signal events with
semi-leptonic decays of b or c hadrons produced in Z decays.

The larger of the jet masses is required to be in the range from $6 \gev{}$ up
to $50 \gev{}$  and the lower one greater than $4 \gev{}$. The upper
mass limit
further removes some semi-leptonic W boson pair decays and the lower
bounds reject two-photon interactions with tau leptons in the final state. 

\begin{figure}[p]
\begin{center}          
\psfig{file=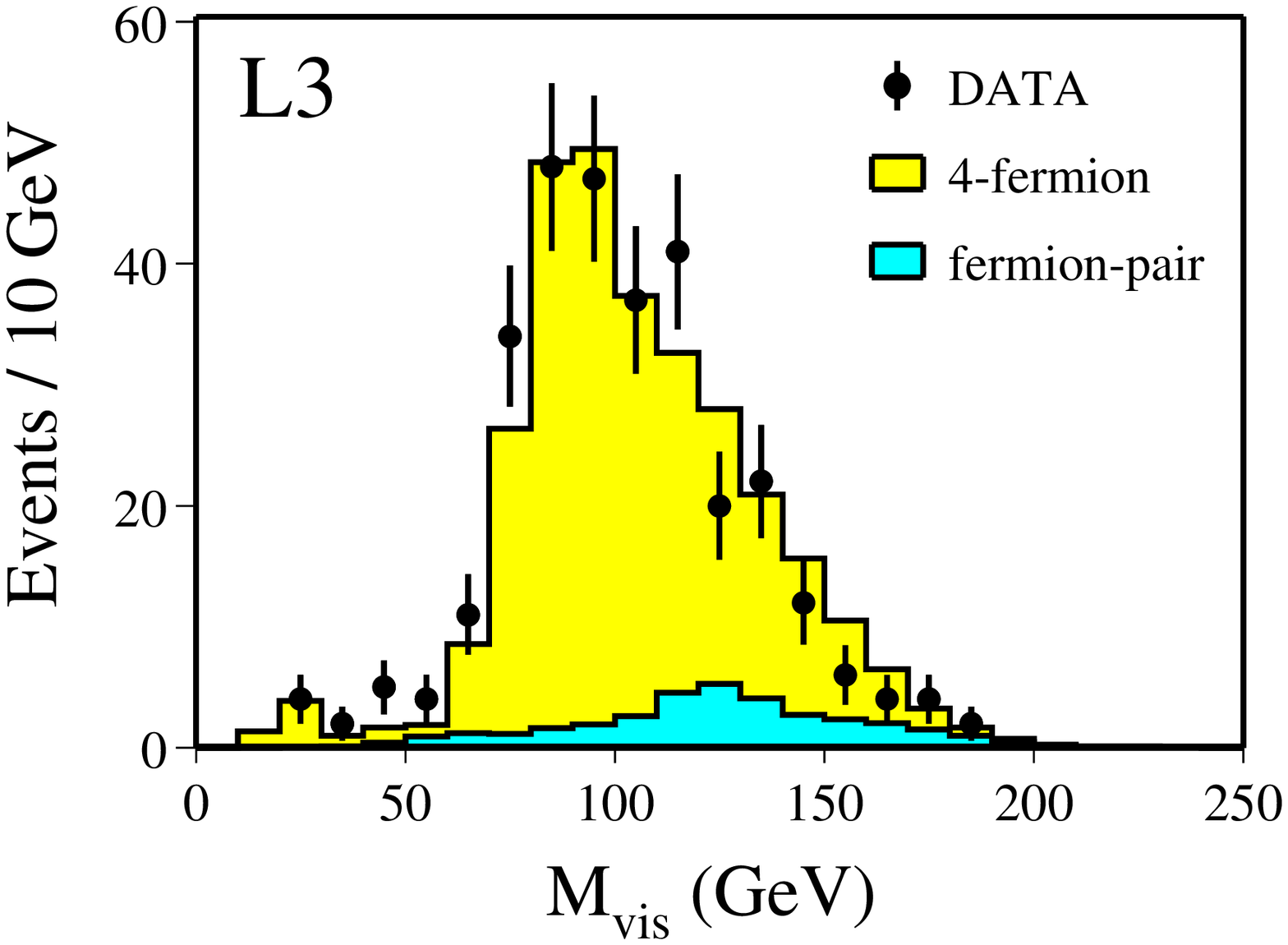,height=12cm}
\icaption{\label{fig_h1} 
The distribution of the visible mass after the hadronic preselection at $\rts=189\gev{}$.}
\end{center}
\end{figure}

Figure~\ref{fig_h1} shows the comparison between data and 
Monte Carlo expectations for the distribution of  the visible mass,
$M_{\rm vis}$ of the preselected events.

\subsection{Heavy Higgs boson selection}

In addition to the preselection described above, the 
correlation between the visible and the
missing mass, $M_{\rm mis}$, is used to select heavy Higgs boson
candidates
in the $\sqrt{s} = 189\gev{}$ data sample.
We define the variable 
$R=(M_{\rm vis}+M_{\rm mis})/(M_{\rm vis}-M_{\rm mis})$
and we require $R < -3.5$ or 
$R > 7$, since the signal has a broad $R$ distribution while for the background
$R$ is close to zero.

A heavy Higgs boson is characterised by relatively low  momentum, hence the
missing momentum of the event should not exceed $40 \gev{}$.
The background due to the
$\QQG$ and two-photon interaction processes is suppressed
by rejecting collinear events and by requiring  a large value of the
event thrust, together with a moderate value of the 
sum of the inter-jet angles, $\Theta_{123}$, when the events
are constrained to have three jets.
In order to reject the residual contributions from the W pair and the
single W processes, an isolation criterion on  the missing momentum
vector is applied. In addition, the upper cut on the maximum value
of the jet masses is tightened to  $30 \gev{}$ while the
minimum has to be less than $20 \gev{}$.

The recoiling mass, 
$M_{\rm h}^{rec}$,
is calculated by constraining the visible mass to the Z boson mass
and imposing energy-momentum conservation~\cite{Acciarri:1997hinv};
its distribution is plotted in Figure~\ref{fig_h2}(a) for the data
and the background. 
With this kinematical constraint, the recoil mass
resolution is $3.5 \gev{}$ in the hypothesis of a Higgs boson mass of
$90\gev{}$. 

After applying the selection described above the dominant process in the
remaining background is the Z boson pair production.

\subsection{Light Higgs boson selection}

The production of a light invisible
Higgs  boson at $\sqrt{s} = 189\gev{}$ is characterised by three main
features, exploited by 
the following selection criteria:
mass of the hadronic system
close to the Z mass $|M_{\rm vis}-M_{\rm Z}|<20 \gev{}$, at least $40\%$ but not more than 60\% of the
centre-of-mass energy visible in the detector and 
missing momentum in the window from  $30 \gev{}$ up to $55 \gev{}$.
This last requirement reduces part of the background arising from
Z boson pair production and
two-photon interactions, the latter being further suppressed
by an upper cut on $\Theta_{123}$. Events from $\QQG$ are rejected by
requiring a large value of the event thrust and the longitudinal momentum
imbalance to be less than $40\%$ of the visible energy.
The residual contribution from W pair production is reduced by a cut
on the threshold $y_{23}$ at which the DURHAM algorithm resolves the event into
three jets from a two-jet topology.

\begin{figure}[p]
\begin{center}          
\psfig{file=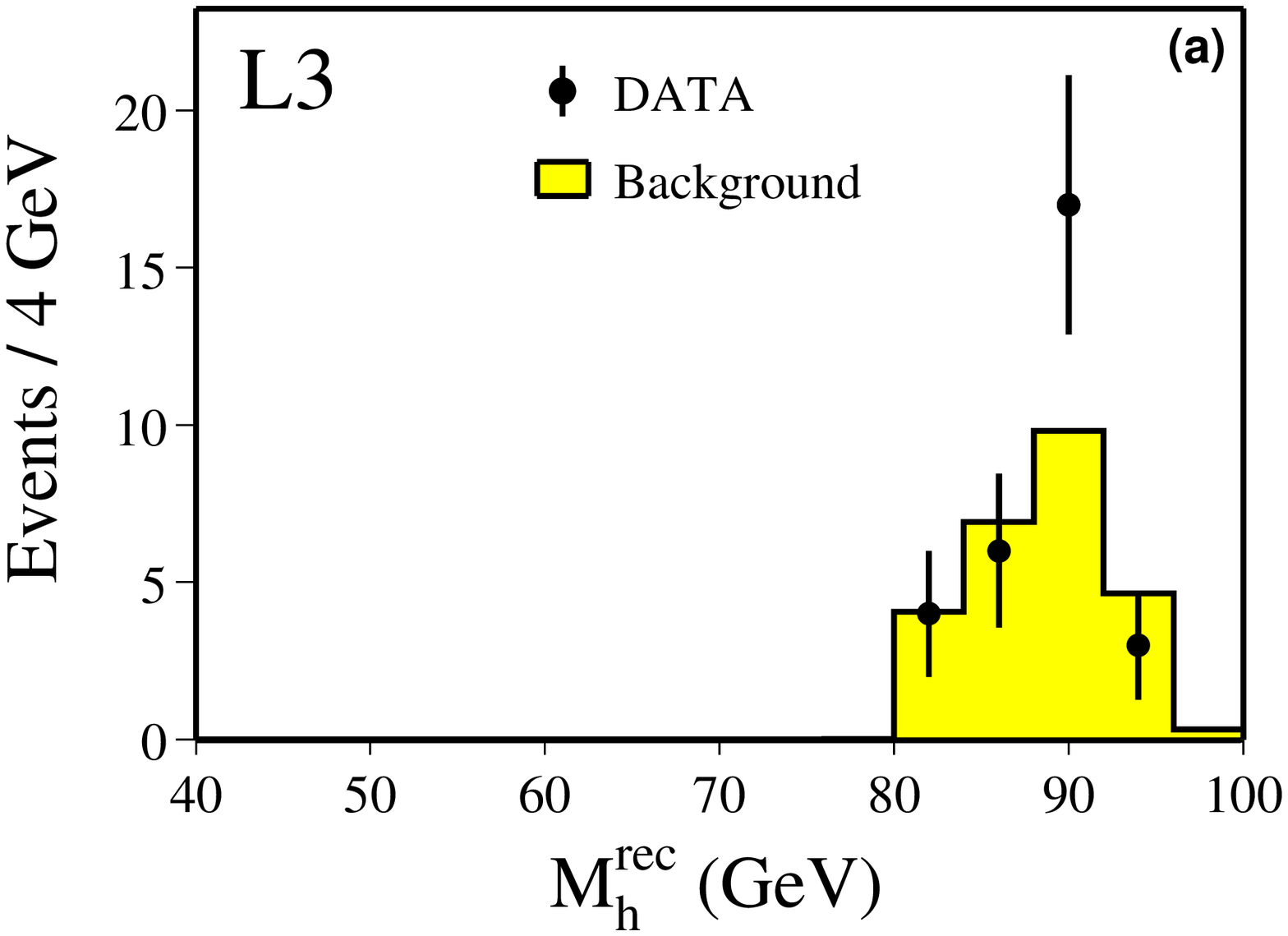,height=7cm}
\psfig{file=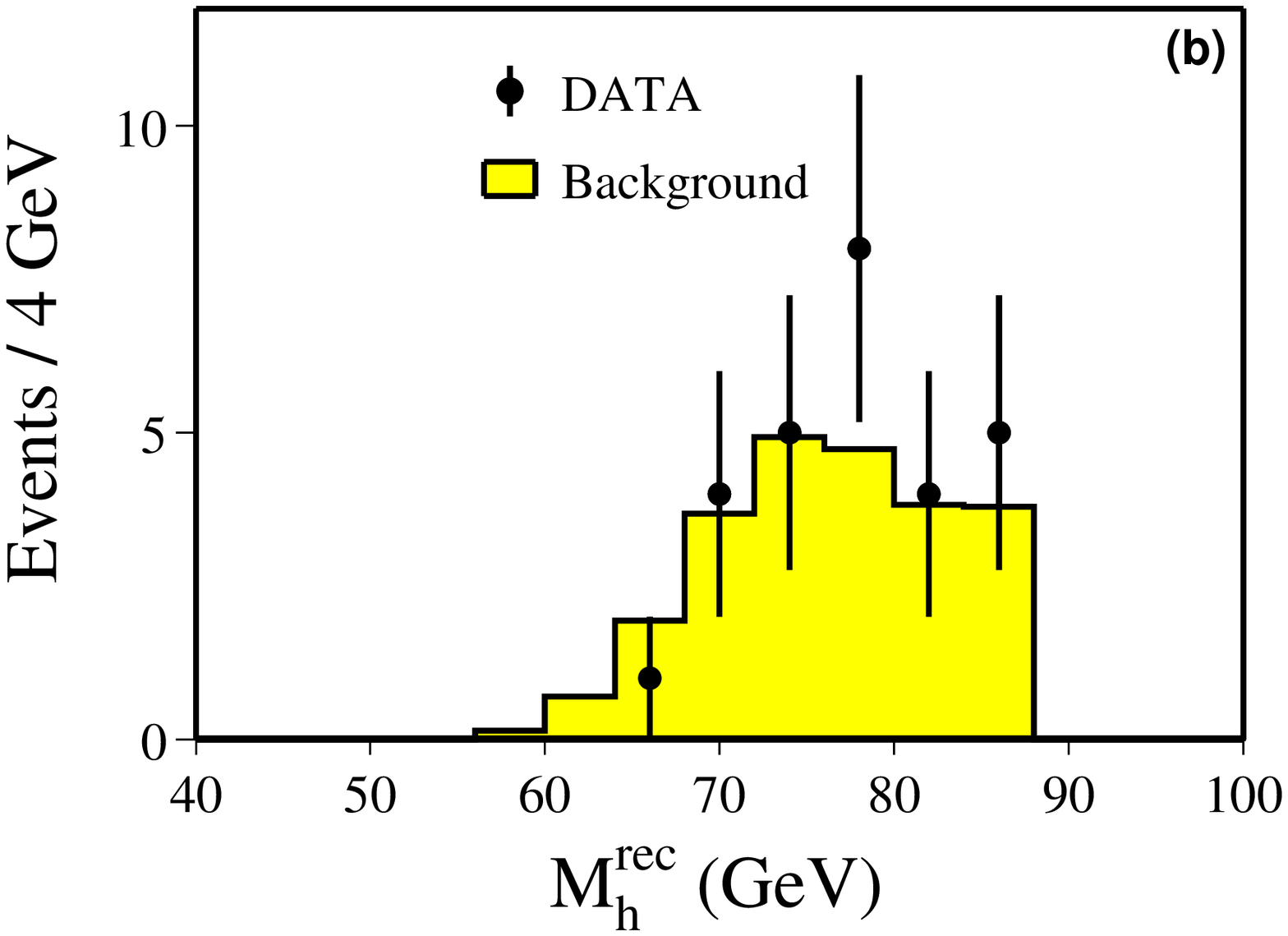,height=7cm}
\psfig{file=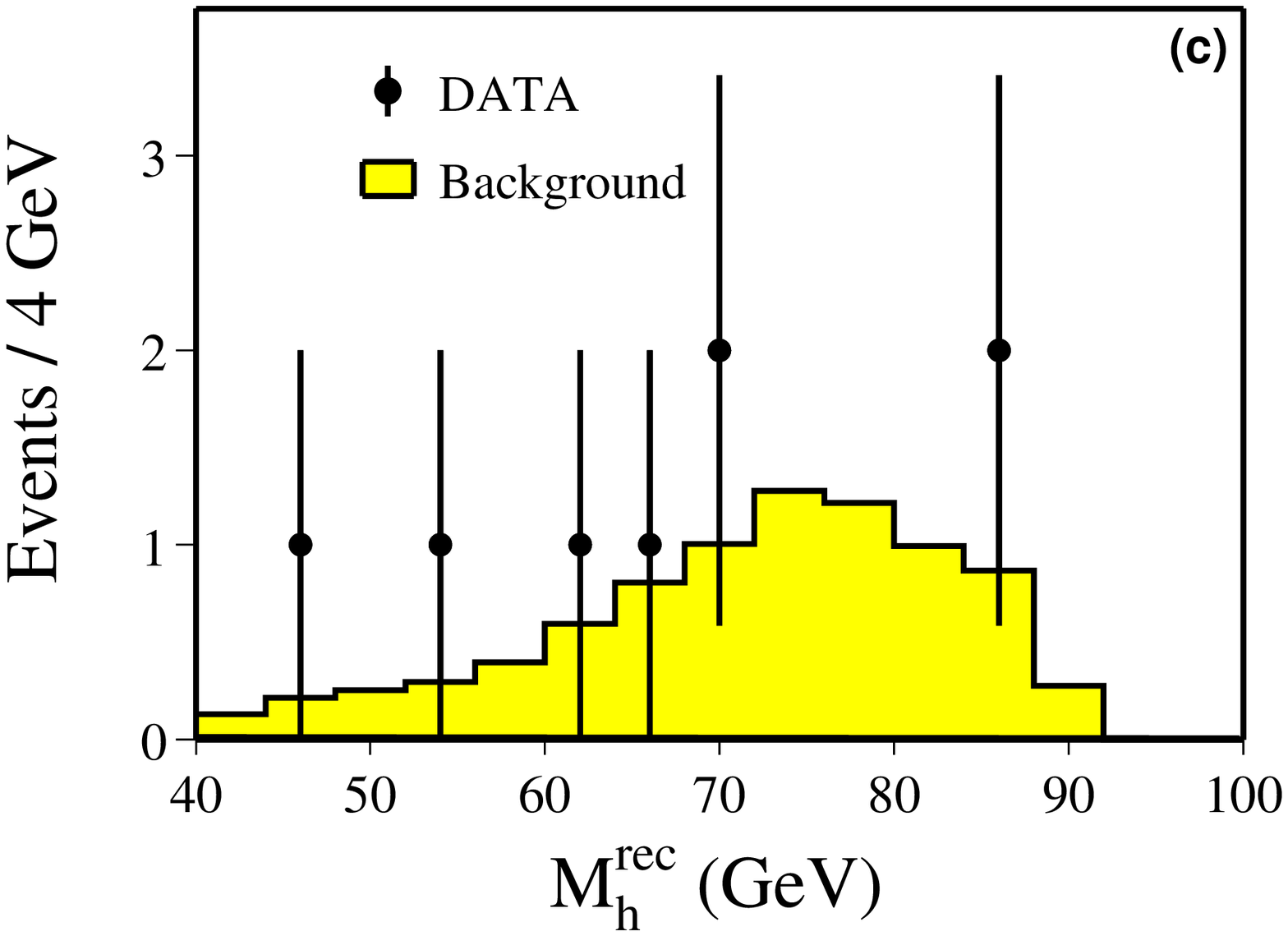,height=7cm}
\icaption{\label{fig_h2} The distribution of the recoil mass (a) after the heavy Higgs boson
selection at $\rts=189\gev{}$, (b) after the light Higgs boson
selection at $\rts=189\gev{}$ and (c) after the final
selection at $\rts=183\gev{}$.}
\end{center}
\end{figure}

The distribution of $M_{\rm h}^{rec}$ for events selected in the data
and the Monte Carlo samples is displayed in
Figure~\ref{fig_h2}(b). 

The selection of the 
hadronic channel at $\rts=183\gev{}$ is similar to this light Higgs 
boson selection. A cut on the transverse momentum imbalance
replaces the cut on the $y_{23}$ parameter and the values of the selection requirements
reflect the different centre-of-mass energies. The final $
M_{\rm h}^{rec}$ spectrum for data and Monte Carlo is shown
in Figure~\ref{fig_h2}(c).
Table~\ref{tab_results} summarises the yields of all the selections
described above.

After applying the two selections described above the dominant process in the
remaining background is the $\WW$ production.

\begin{table} [ht]
\begin{center}
\begin{tabular}{|c|c|c|c|c|c|c|}
\hline
&\multicolumn{3}{|c|}{$\sqrt{s} = 189 \gev{}$} & $\sqrt{s} =  183 \gev{}$ \\
              & Preselection & Heavy Higgs boson & Light Higgs boson& Final Selection\\
\hline
Data          &   304\phantom{.}\phantom{0}   &  30\phantom{.}\phantom{0}     &  27\phantom{.}\phantom{0}   & 8\phantom{.}\phantom{0}    \\
\hline
Background MC                         &   300.3 & 25.8    &  23.6 & 8.5   \\
\hline\hline
$\epsilon\ (\%)$ ($M_{\rm h}=65 \gev{}$)     &   \phantom{0}53.1  & \phantom{0}3.0     & 19.2  & 22.0  \\
$\epsilon\ (\%)$ ($M_{\rm h}=85 \gev{}$)     &   \phantom{0}54.0  & 30.2    & 20.0  & 28.0\\
$\epsilon\ (\%)$ ($M_{\rm h}=90 \gev{}$)     &   \phantom{0}54.1  &  37.8   &  \phantom{0}7.6 & 18.5 \\
$\epsilon\ (\%)$ ($M_{\rm h}=95 \gev{}$)     &   \phantom{0}46.6  & 32.2    & \phantom{0}2.1   & -- \\
\hline
\end{tabular}       
\icaption{\label{tab_results}
Number of events expected from Standard Model processes compared to the number
of  data events  selected by the hadronic selections.
The signal efficiencies ($\epsilon$) for several Higgs boson masses $M_{\rm h}$
are also shown.}
\end{center}
\end{table}


%
\section{Search in the leptonic channels}

\begin{figure}[p]
\begin{center}          
\hspace*{-0.5cm}
\psfig{file=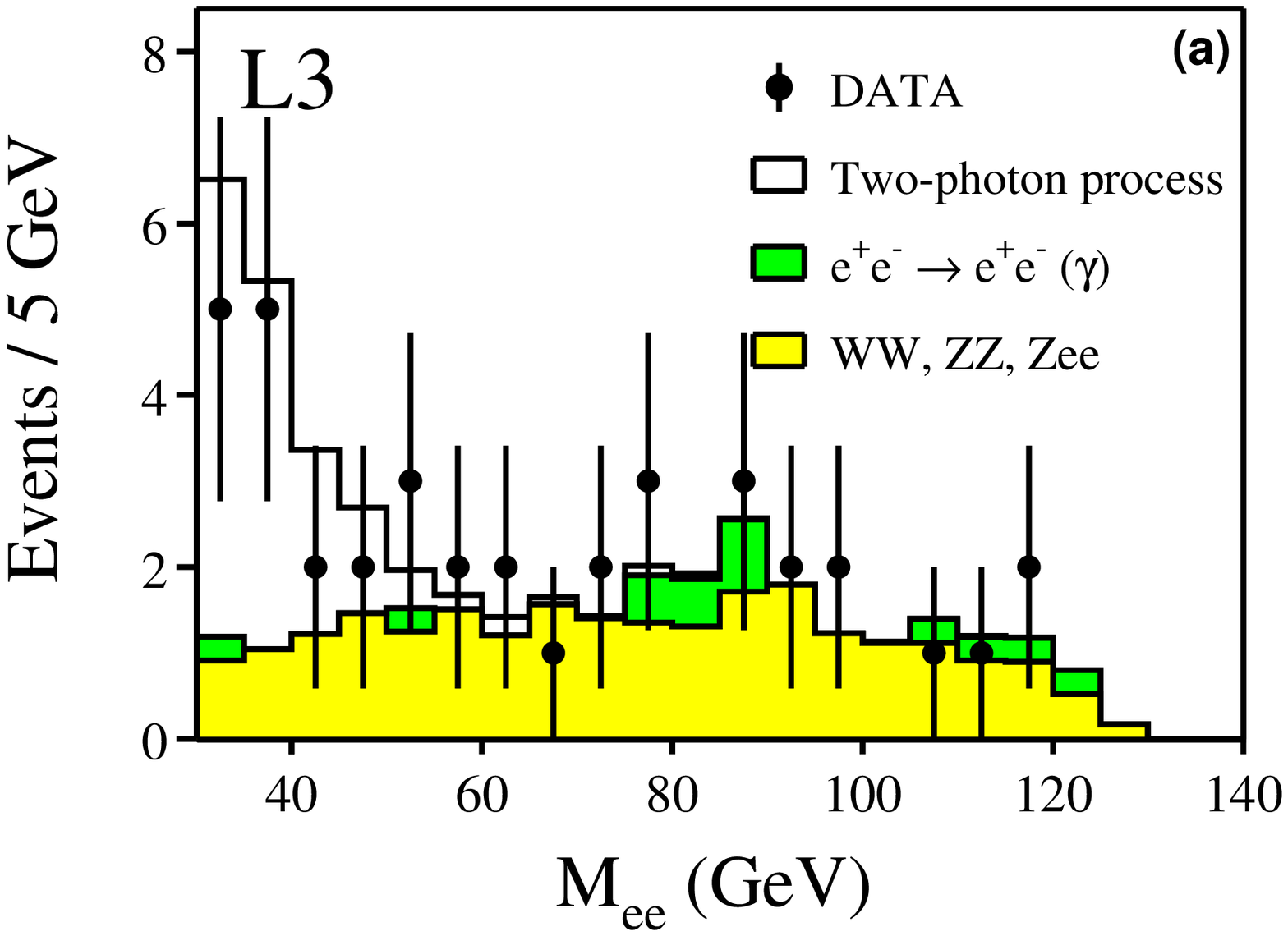,height=10cm} 
\psfig{file=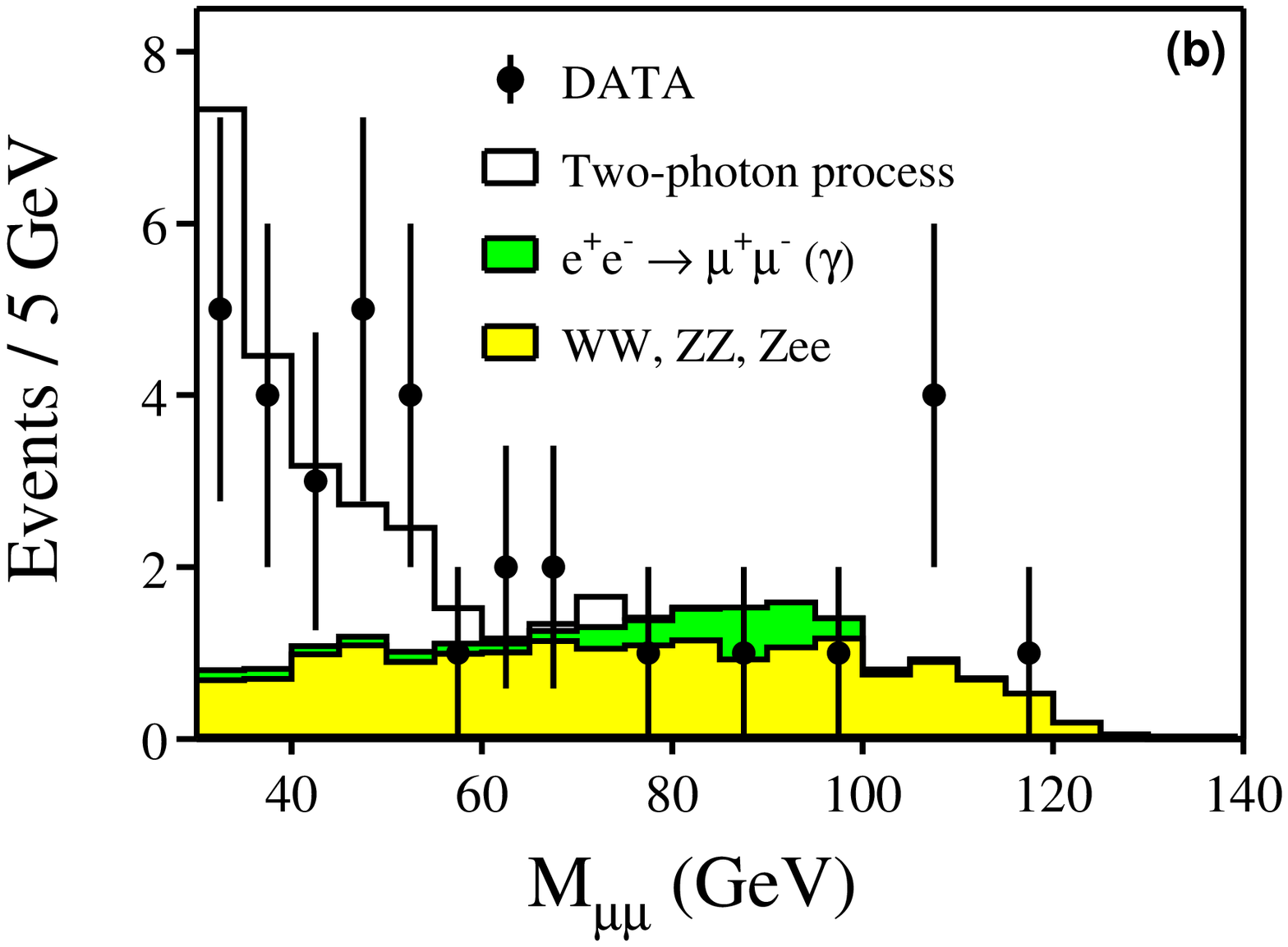,height=10cm}
\icaption{\label{fig_ll_pre1}
Distribution of (a) the dielectron and (b) the dimuon invariant mass at $\sqrt{s}= 189 \gev{}$  
after the preselection is applied.}
\end{center}
\end{figure}

The search for an invisibly decaying Higgs boson produced in association
with a Z boson decaying into leptons is designed to be almost independent of the Higgs boson
mass in the investigated range. Low multiplicity events with
a pair of high energy muons or electrons are selected.
These are separated from  fermion pair
production events by requiring large acoplanarity  and  visible
energy between $5\%$ and $70\%$ of the centre-of-mass energy.
The lepton energy has  to be less than $90\%$ of the beam energy
to further reject Bhabha scattering events.
Two-photon interactions are suppressed by requiring the lepton
pair invariant mass to be larger than $30\GeV{}$ and
low energy depositions in the forward calorimeters.
Events with muons should have at least one scintillator
in time with the beam crossing in order
to remove cosmic-ray background.   
The yield of this preselection is presented in Table~\ref{tab_ll_sel}, while
Figure~\ref{fig_ll_pre1} displays the spectra of the 
lepton pair invariant mass, $M_{\ell\ell}$,
for data and Standard Model Monte Carlo events.

\begin{table} [h]
\begin{center}
\begin{tabular}{|c|c|c||c|c|    }
\hline
 & \multicolumn{2}{|c||}{Electrons} & \multicolumn{2}{c|}{Muons} \\ 
\hline \hline
  & Preselection & Final selection & Preselection & Final selection\\ 
\hline 
Data & 38\phantom{.0}      &  \phantom{0}2\phantom{.0}  & 34\phantom{.0}   & 2\phantom{.0}   \\ 
\hline   
Background MC   & 41.4 &  \phantom{0}2.2 & 36.5 &  \phantom{0}1.6    \\ 
\hline
$\epsilon\ (\%)$ ($M_{\rm h}=65 \gev{}$) & 52.9 & 36.5 & 42.1 &  19.1\\ 
$\epsilon\ (\%)$ ($M_{\rm h}=85 \gev{}$) & 55.4 & 41.3 & 42.3 &  20.9\\  
$\epsilon\ (\%)$ ($M_{\rm h}=90 \gev{}$) & 55.4 & 39.6 & 45.8 &  25.1\\
$\epsilon\ (\%)$ ($M_{\rm h}=95 \gev{}$) & 55.3 & 42.1 & 47.7 &  30.2\\ 
\hline 
\end{tabular} 
\icaption{\label{tab_ll_sel} Number of 
events observed and expected from Standard Model processes at $\rts =
189 \gev{}$ after the preselections and the final selections.
Signal efficiencies ($\epsilon$) for different Higgs mass hypotheses are also shown.
The background to the final selected sample is composed of one third
Z boson-pair events and two thirds W boson-pair events.}
\end{center}
\end{table}

Residual events due to the radiative return to the Z resonance where the
photon remains undetected in the beam pipe are rejected by requiring the missing
momentum to point away from the beam axis. Tau pair production can
yield acoplanar lepton pairs that satisfy the selection criteria
described above. 
In the hypothesis that the lepton pair originates from a single
particle, we require the cosine of the most energetic lepton 
emission angle $\theta^*$ in the Z boson rest frame not to exceed 0.95.

The contribution from  two-photon interactions is eliminated by
tightening the cut on the lepton invariant masses, 
$70 \gev{} <M_{\ell\ell}<\ 110 \gev{}$; this is also
effective against a significant portion of the fully leptonic decays
of W bosons. Final states with an electron or muon
pair and  two neutrinos, produced by Z boson pairs, constitute an
irreducible background but their cross section is relatively low.

\begin{figure}[p]
\begin{center}          
\hspace*{-0.5cm}
\psfig{file=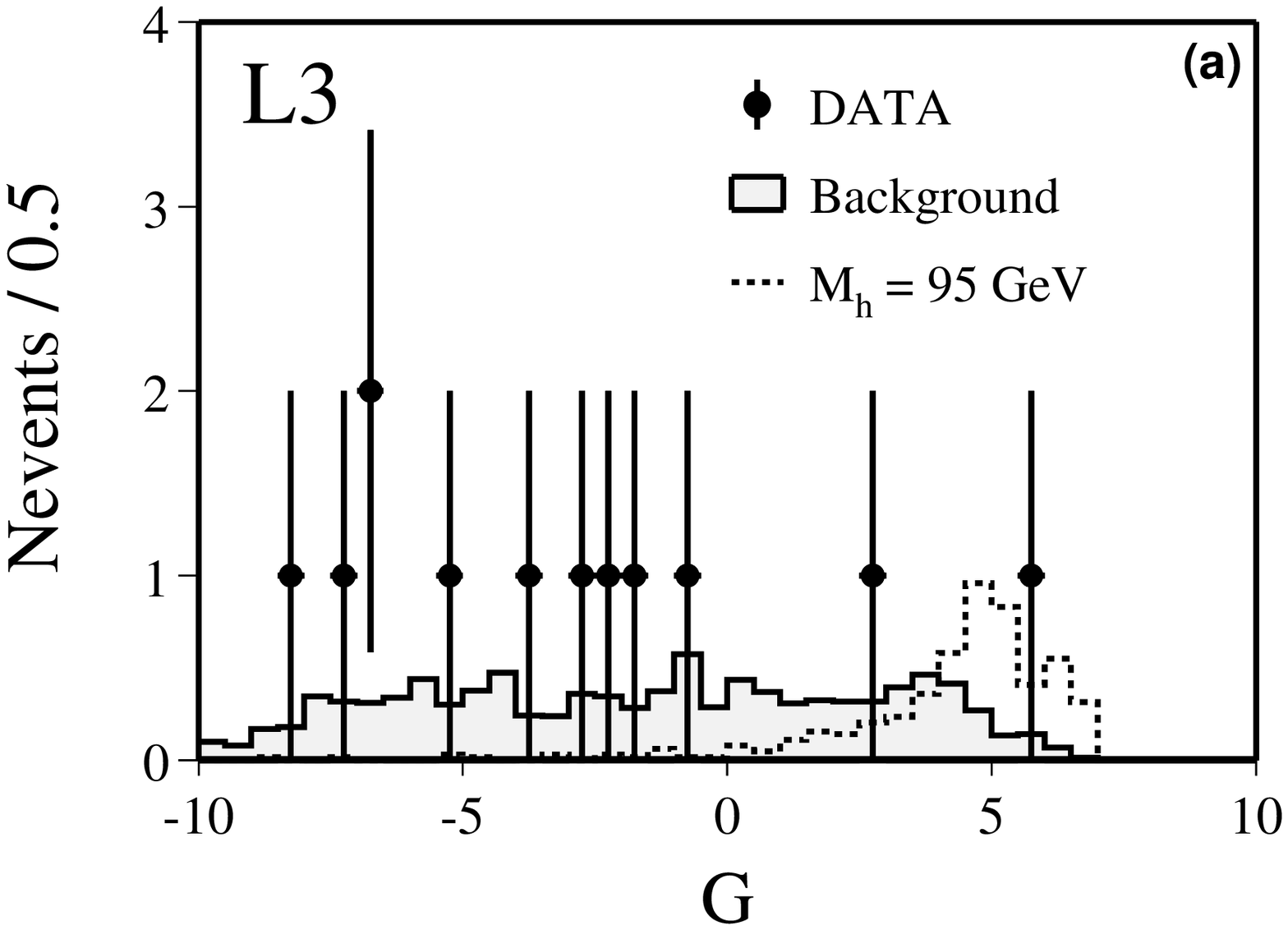,height=10cm} 
\psfig{file=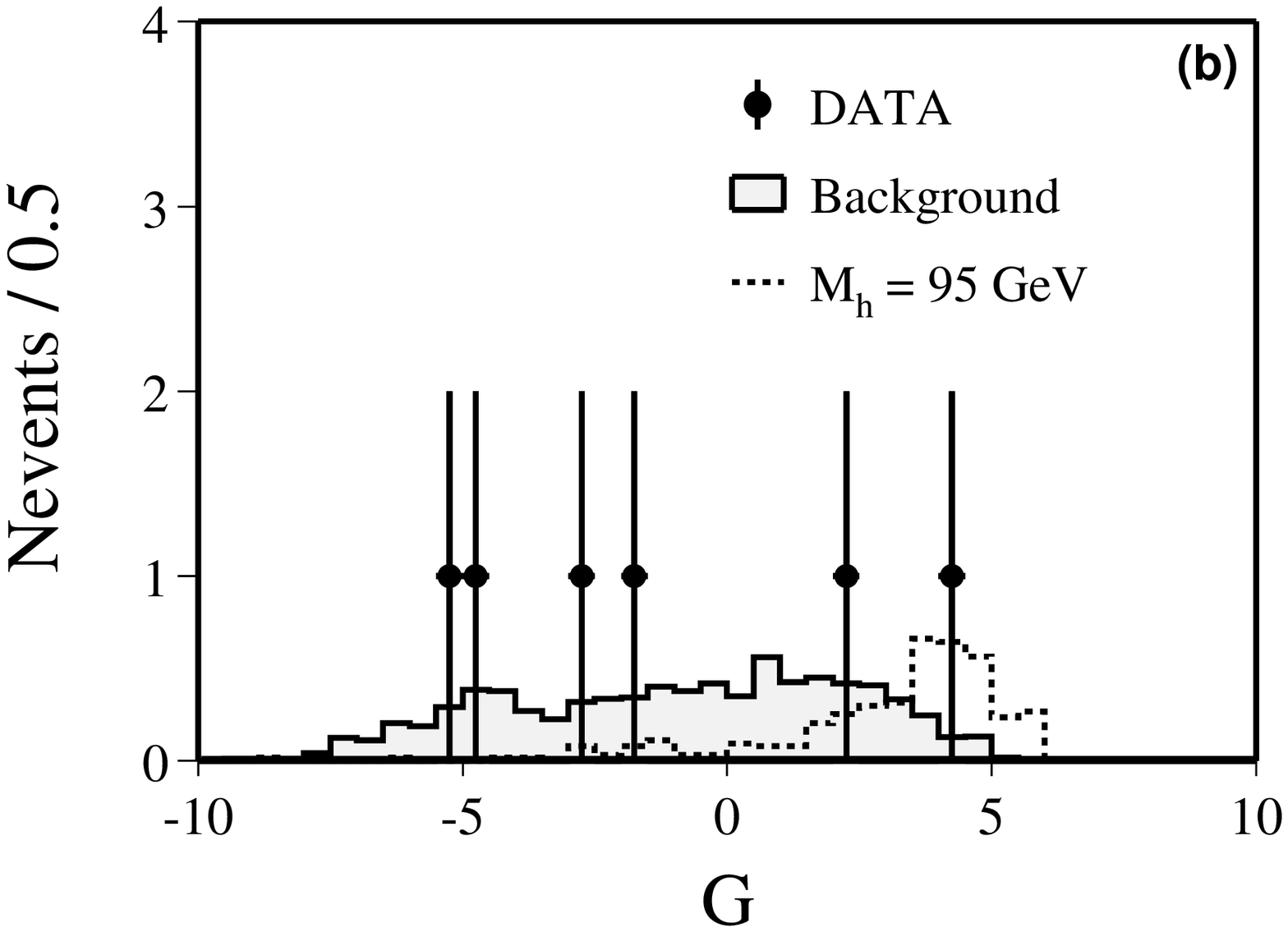,height=10cm}
\icaption{\label{fig_this_disc} 
Distributions of the final likelihood variable, $G$, (a) for the electron
and (b) for the muon selections at $\rts=189 \gev{}$,
for data and the expected  background. A possible Higgs signal
($M_{\rm h}=95 \gev{}$) with an arbitrary cross section is also
shown.}
\end{center}
\end{figure}

The visible energy $E_{\rm vis}$, $\cos{~\theta^*}$, $M_{\ell\ell}$
and the velocity $\beta$ of the dilepton system are
combined into a single likelihood variable $G$, defined as:

\begin{center}
$G=\sum_i \log{(P_S^{i}(x))}-\log{(P_B^{i}(x))}$.
\end{center}
The index $i$ runs over the four variables and $P_S^{i}(x)$ and $P_B^{i}(x)$
are the probability densities for the $i$-th variable  to have a value $x$
in the signal or background hypotheses, respectively. 
These densities are calculated for each event by interpolating between
the two signal Monte Carlo 
samples whose generated Higgs masses are closer to the event
missing mass which is taken as the Higgs boson mass hypothesis.
The Z boson pair  background is not included in this calculation.
Figure~\ref{fig_this_disc} shows 
distributions of $G$ for the data and the expected  Monte Carlo
background and signal for a 
Higgs boson mass of $95 \gev{}$. 

The number of selected events and the signal efficiency after the
optimization~\cite{susy} of a cut on $G$ are reported in Table~\ref{tab_ll_sel}.
The observed resolution on the missing mass is $1.1 \gev{}$ 
in the electron channel, and $5.1 \gev{}$ in the muon channel for a
Higgs boson mass of $90 \gev{}$. 

\begin{figure}[p]
\begin{center}          
\hspace*{-0.5cm}
\psfig{file=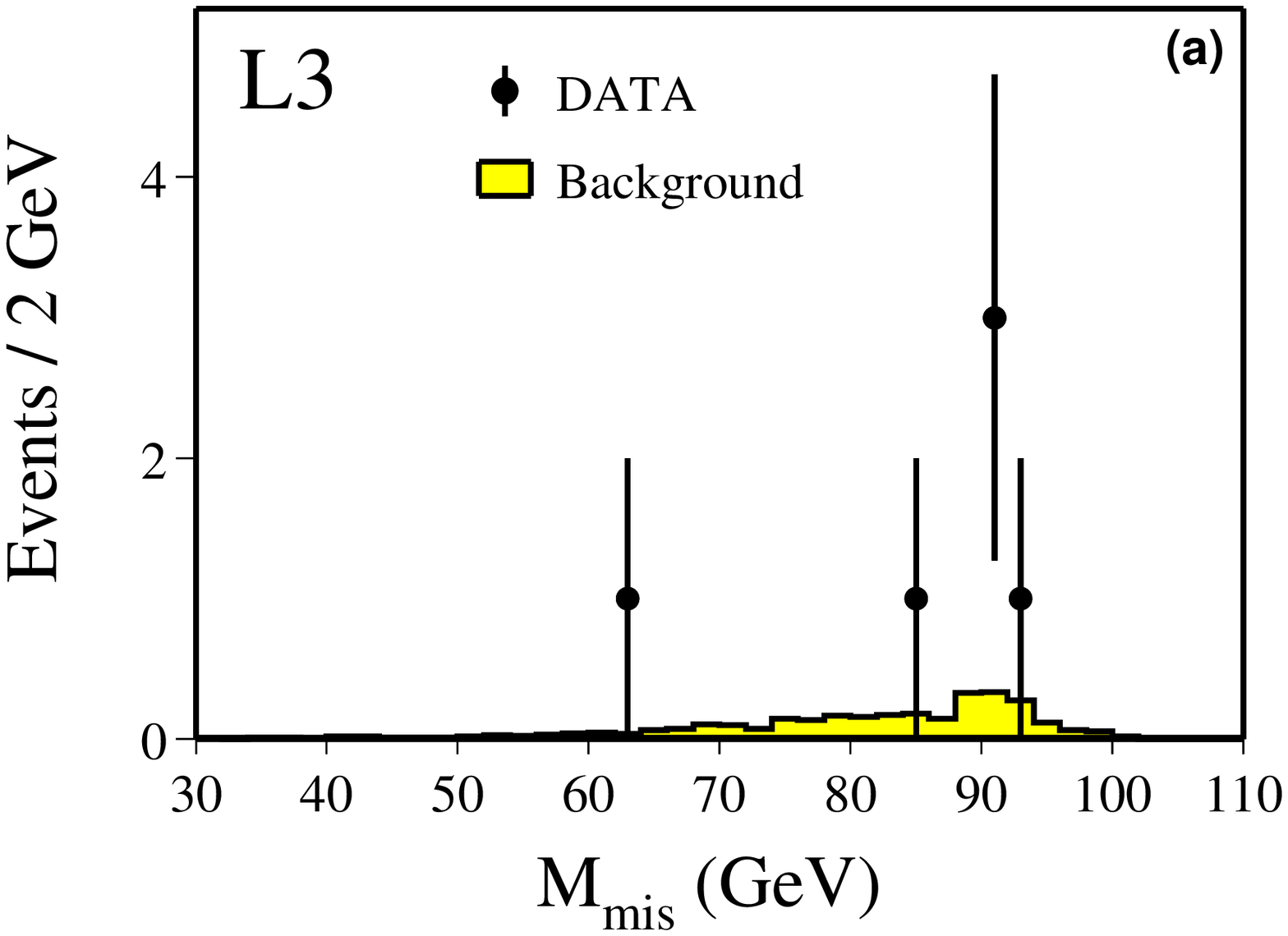,height=10cm} 
\psfig{file=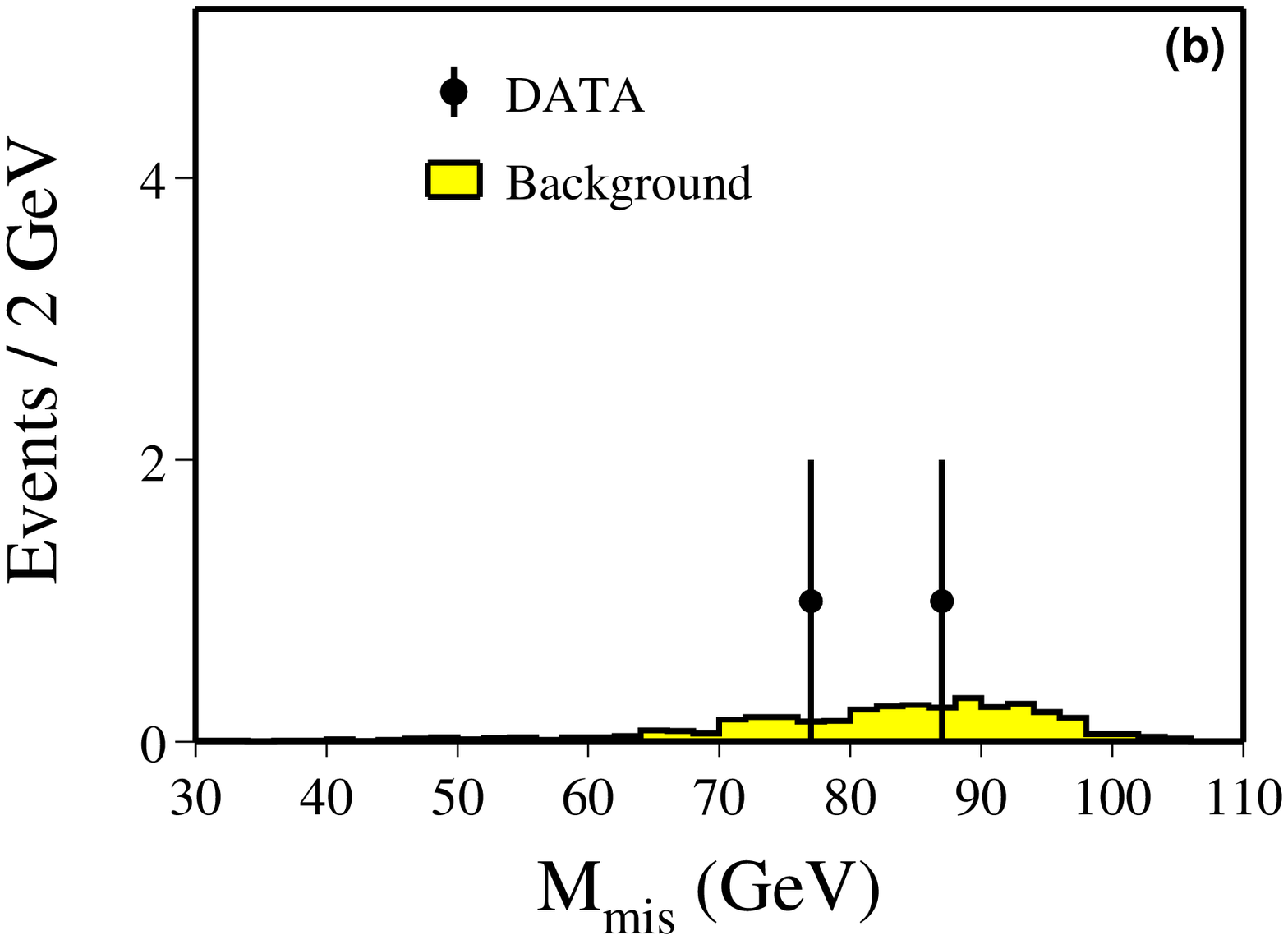,height=10cm}
\icaption{\label{spectr_l}
The missing mass distributions (a) in the electron channel
and (b) in the muon channel for the combined $\rts =183 \gev{}$ and $189 \gev{}$ data
samples.}
\end{center}
\end{figure}

A cut-based analysis is developed for the  $\rts = 183 \gev{}$ data
sample making use of the following selection criteria: 
$30 \gev{}<E_{\rm vis}<120 \gev{}$, $\cos{~\theta^*}<0.95$, $80 \gev{}<M_{\ell\ell}<100 \gev{}$
and $0.05<\beta<0.55$.
In the electron channel, the signal efficiency is $45\%$ and
4 events are observed for 1.4 expected background events.
In the muon channel, the signal efficiency is  $28\%$ and 
no events are observed while 1.7 background events are expected.
Figure~\ref{spectr_l} displays the missing mass distributions for
the search in the leptonic channel
for the combined  $\rts = 183 \gev{}$ and  $\rts = 189 \gev{}$ samples.

\section{Systematic uncertainties} 

Two sources of systematic uncertainties, summarised in Table~\ref{tab_syst},
can affect the results.
The first is the limited amount of Monte Carlo statistics, which gives 
the systematic errors on the signal and background efficiencies listed as ``MC Stat.'' in
Table~\ref{tab_syst}. The second is the quality of the Monte Carlo
description of the background processes. This is studied using data and Monte
Carlo samples containing essentially $\WW$ and ZZ background events. These samples contain about 1100 events for the hadronic
channel and 500 for the leptonic ones. 
The data  distributions in these new samples of each selection
variable $i$, except the
likelihood $G$, are compared with those of the Monte Carlo, determining 
their systematic shifts $s_i$ and the corresponding statistical
errors $\sigma_i$. 
All the selection cuts are then shifted by $s_i \pm
\sigma_i$, where the sign of $\sigma_i$ is chosen so as to obtain the lowest
efficiency for the single cut on the variable $i$.
The difference between the efficiency of the selection using the
shifted cuts and that of the nominal one is taken as the systematic uncertainty.
These errors are summarised as ``Syst.'' in Table~\ref{tab_syst} and are summed
in quadrature with the Monte Carlo statistical uncertainties to obtain
the total systematic uncertainty, listed as ``Total'' in Table~\ref{tab_syst}.

\begin{table} [h]
\begin{center}

\begin{tabular}{|c|c|c|c||c|c|c|}
\hline
                   & \multicolumn{3}{|c||}{Background} & \multicolumn{3}{c|}{Signal} \\ 
\hline
                   &  MC Stat.   &   Syst.   & Total &   MC Stat.   &    Syst.   & Total\\
\hline
Heavy hadronic &    1.0      &    5.0     &        5.0            &    3.0       &     2.5       &        4.0       \\
\hline
Light hadronic &    1.0      &    5.0     &        5.0            &   4.0        &     5.5       &        7.0       \\
\hline
Electrons       &    6.0       &   1.5     &        6.0            &     5.0    &     2.0     &        5.5       \\
\hline
Muons      &    6.5      &   5.0      &        8.0            &     6.0    &    3.5     &       7.0       \\
\hline 
\end{tabular} 
\icaption{\label{tab_syst}Relative systematic uncertainties in percent on the
signal and background efficiencies for each analysis.}
\end{center}
\end{table}

\section{Results} 

\begin{figure}[htbp]
\begin{center}
\psfig{file=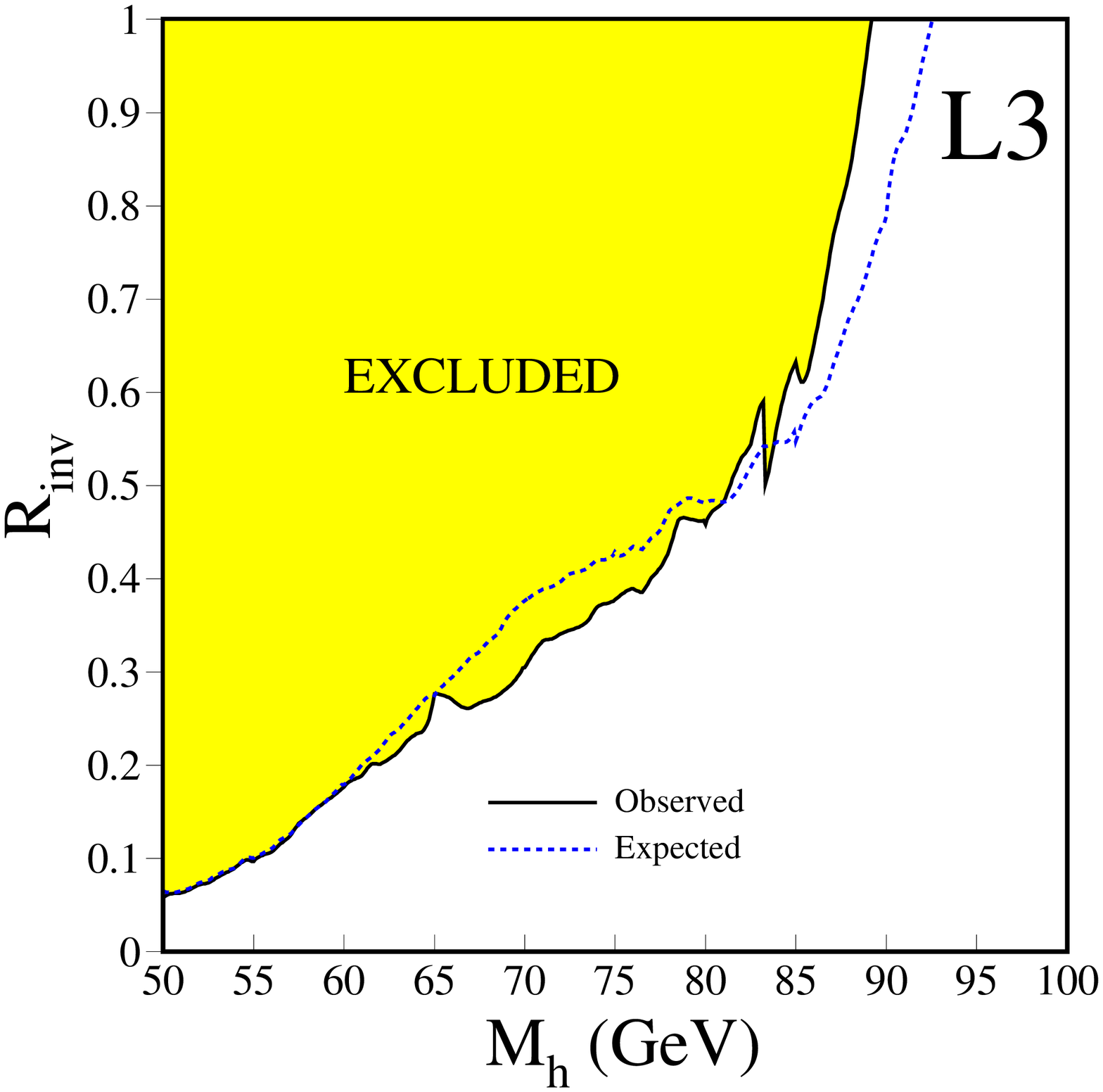,height=15 cm}
\vskip 1.15 cm
\icaption{\label{exclusion}
Observed and expected upper limits on the ratio of
the invisibly-decaying Higgs boson cross section to that of the
Standard Model Higgs boson, as function of the Higgs boson mass.
The shaded area is excluded at least at $95\%$ confidence level.}
\end{center}
\end{figure}

No indication of the production of a Higgs boson with invisible decays is
found. As both the production cross section and the
branching ratios are model dependent, it is useful to
introduce the ratio $R_{\rm inv}=\mathrm{BR}(\mathrm{h} \to
\mathrm{invisible\ particles})\times\sigma(\EE \to \mathrm{hZ})/
\sigma(\EE \to \mathrm{H_{SM}Z})$, where $\mathrm{H_{SM}}$
is the Standard Model Higgs boson. A limit on $R_{\rm inv}$ is
calculated~\cite{freq_cl} as a function of the Higgs boson mass making
use of  the  mass distributions presented
in Figures~\ref{fig_h2} and~\ref{spectr_l}.
In the determination of the limit 
the Standard Model Higgs boson cross section as given by the HZHA
generator \cite{hzha} is used and the  signal and background
efficiencies are lowered by their systematic uncertainties. Results 
obtained at lower energies~\cite{Acciarri:1997hinv} are included.
Figure~\ref{exclusion} shows the $95\%$ confidence level (CL) upper limit on 
$R_{\rm inv}$ as a function of the  Higgs mass $M_{\rm h}$. For the
value of $R_{\rm inv}=1$ the  $95\%$ CL lower limit on the
Higgs boson mass is:
\begin{displaymath}
 M_{\rm h}>89.2 \gev{}.
\end{displaymath}
The expected lower limit is $92.6 \gev{}$.
\section*{Acknowledgements}

We wish to express our gratitude to the CERN Accelerator Divisions for the good performance
of the LEP machine. We acknowledge the efforts of the engineers and technicians who have
participated in the construction and maintenance of this experiment.

%
%
\newpage
\section*{Author List}
\typeout{   }     
\typeout{Using author list for paper 206 -?}
\typeout{$Modified: Mon Feb 14 17:42:04 2000 by clare $}
\typeout{!!!!  This should only be used with document option a4p!!!!}
\typeout{   }
%
%
%
%
%
%

\newcount\tutecount  \tutecount=0
\def\tutenum#1{\global\advance\tutecount by 1 \xdef#1{\the\tutecount}}
\def\tute#1{$^{#1}$}
\tutenum\aachen            
\tutenum\nikhef            
\tutenum\mich              
\tutenum\lapp              
\tutenum\basel             
\tutenum\lsu               
\tutenum\beijing           
\tutenum\berlin            
\tutenum\bologna           
\tutenum\tata              
\tutenum\ne                
\tutenum\bucharest         
\tutenum\budapest          
\tutenum\mit               
\tutenum\debrecen          
\tutenum\florence          
\tutenum\cern              
\tutenum\wl                
\tutenum\geneva            
\tutenum\hefei             
\tutenum\seft              
\tutenum\lausanne          
\tutenum\lecce             
\tutenum\lyon              
\tutenum\madrid            
\tutenum\milan             
\tutenum\moscow            
\tutenum\naples            
\tutenum\cyprus            
\tutenum\nymegen           
\tutenum\caltech           
\tutenum\perugia           
\tutenum\cmu               
\tutenum\prince            
\tutenum\rome              
\tutenum\peters            
\tutenum\potenza           
\tutenum\salerno           
\tutenum\ucsd              
\tutenum\santiago          
\tutenum\sofia             
\tutenum\korea             
\tutenum\alabama           
\tutenum\utrecht           
\tutenum\purdue            
\tutenum\psinst            
\tutenum\zeuthen           
\tutenum\eth               
\tutenum\hamburg           
\tutenum\taiwan            
\tutenum\tsinghua          
{
\parskip=0pt
\noindent
{\bf The L3 Collaboration:}
\ifx\selectfont\undefined
 \baselineskip=10.8pt
 \baselineskip\baselinestretch\baselineskip
 \normalbaselineskip\baselineskip
 \ixpt
\else
 \fontsize{9}{10.8pt}\selectfont
\fi
\medskip
\tolerance=10000
\hbadness=5000
\raggedright
\hsize=162truemm\hoffset=0mm
\def\r{\rlap,}
\noindent

M.Acciarri\r\tute\milan\
P.Achard\r\tute\geneva\ 
O.Adriani\r\tute{\florence}\ 
M.Aguilar-Benitez\r\tute\madrid\ 
J.Alcaraz\r\tute\madrid\ 
G.Alemanni\r\tute\lausanne\
J.Allaby\r\tute\cern\
A.Aloisio\r\tute\naples\ 
M.G.Alviggi\r\tute\naples\
G.Ambrosi\r\tute\geneva\
% ...
\newpage
\begin{list}{A}{\itemsep=0pt plus 0pt minus 0pt\parsep=0pt plus 0pt minus 0pt
                \topsep=0pt plus 0pt minus 0pt}
\item[\aachen]
 I. Physikalisches Institut, RWTH, D-52056 Aachen, FRG$^{\S}$\\
 III. Physikalisches Institut, RWTH, D-52056 Aachen, FRG$^{\S}$
\item[\nikhef] National Institute for High Energy Physics, NIKHEF, 
     and University of Amsterdam, NL-1009 DB Amsterdam, The Netherlands
\item[\mich] University of Michigan, Ann Arbor, MI 48109, USA
\item[\lapp] Laboratoire d'Annecy-le-Vieux de Physique des Particules, 
     LAPP,IN2P3-CNRS, BP 110, F-74941 Annecy-le-Vieux CEDEX, France
\item[\basel] Institute of Physics, University of Basel, CH-4056 Basel,
     Switzerland
\item[\lsu] Louisiana State University, Baton Rouge, LA 70803, USA
\item[\beijing] Institute of High Energy Physics, IHEP, 
  100039 Beijing, China$^{\triangle}$ 
\item[\berlin] Humboldt University, D-10099 Berlin, FRG$^{\S}$
\item[\bologna] University of Bologna and INFN-Sezione di Bologna, 
     I-40126 Bologna, Italy
\item[\tata] Tata Institute of Fundamental Research, Bombay 400 005, India
\item[\ne] Northeastern University, Boston, MA 02115, USA
\item[\bucharest] Institute of Atomic Physics and University of Bucharest,
     R-76900 Bucharest, Romania
\item[\budapest] Central Research Institute for Physics of the 
     Hungarian Academy of Sciences, H-1525 Budapest 114, Hungary$^{\ddag}$
\item[\mit] Massachusetts Institute of Technology, Cambridge, MA 02139, USA
\item[\debrecen] KLTE-ATOMKI, H-4010 Debrecen, Hungary$^\P$
\item[\florence] INFN Sezione di Firenze and University of Florence, 
     I-50125 Florence, Italy
\item[\cern] European Laboratory for Particle Physics, CERN, 
     CH-1211 Geneva 23, Switzerland
\item[\wl] World Laboratory, FBLJA  Project, CH-1211 Geneva 23, Switzerland
\item[\geneva] University of Geneva, CH-1211 Geneva 4, Switzerland
\item[\hefei] Chinese University of Science and Technology, USTC,
      Hefei, Anhui 230 029, China$^{\triangle}$
\item[\seft] SEFT, Research Institute for High Energy Physics, P.O. Box 9,
      SF-00014 Helsinki, Finland
\item[\lausanne] University of Lausanne, CH-1015 Lausanne, Switzerland
\item[\lecce] INFN-Sezione di Lecce and Universit\'a Degli Studi di Lecce,
     I-73100 Lecce, Italy
\item[\lyon] Institut de Physique Nucl\'eaire de Lyon, 
     IN2P3-CNRS,Universit\'e Claude Bernard, 
     F-69622 Villeurbanne, France
\item[\madrid] Centro de Investigaciones Energ{\'e}ticas, 
     Medioambientales y Tecnolog{\'\i}cas, CIEMAT, E-28040 Madrid,
     Spain${\flat}$ 
\item[\milan] INFN-Sezione di Milano, I-20133 Milan, Italy
\item[\moscow] Institute of Theoretical and Experimental Physics, ITEP, 
     Moscow, Russia
\item[\naples] INFN-Sezione di Napoli and University of Naples, 
     I-80125 Naples, Italy
\item[\cyprus] Department of Natural Sciences, University of Cyprus,
     Nicosia, Cyprus
\item[\nymegen] University of Nijmegen and NIKHEF, 
     NL-6525 ED Nijmegen, The Netherlands
\item[\caltech] California Institute of Technology, Pasadena, CA 91125, USA
\item[\perugia] INFN-Sezione di Perugia and Universit\'a Degli 
     Studi di Perugia, I-06100 Perugia, Italy   
\item[\cmu] Carnegie Mellon University, Pittsburgh, PA 15213, USA
\item[\prince] Princeton University, Princeton, NJ 08544, USA
\item[\rome] INFN-Sezione di Roma and University of Rome, ``La Sapienza",
     I-00185 Rome, Italy
\item[\peters] Nuclear Physics Institute, St. Petersburg, Russia
\item[\potenza] INFN-Sezione di Napoli and University of Potenza, 
     I-85100 Potenza, Italy
\item[\salerno] University and INFN, Salerno, I-84100 Salerno, Italy
\item[\ucsd] University of California, San Diego, CA 92093, USA
\item[\santiago] Dept. de Fisica de Particulas Elementales, Univ. de Santiago,
     E-15706 Santiago de Compostela, Spain
\item[\sofia] Bulgarian Academy of Sciences, Central Lab.~of 
     Mechatronics and Instrumentation, BU-1113 Sofia, Bulgaria
\item[\korea]  Laboratory of High Energy Physics, 
     Kyungpook National University, 702-701 Taegu, Republic of Korea
\item[\alabama] University of Alabama, Tuscaloosa, AL 35486, USA
\item[\utrecht] Utrecht University and NIKHEF, NL-3584 CB Utrecht, 
     The Netherlands
\item[\purdue] Purdue University, West Lafayette, IN 47907, USA
\item[\psinst] Paul Scherrer Institut, PSI, CH-5232 Villigen, Switzerland
\item[\zeuthen] DESY, D-15738 Zeuthen, 
     FRG
\item[\eth] Eidgen\"ossische Technische Hochschule, ETH Z\"urich,
     CH-8093 Z\"urich, Switzerland
\item[\hamburg] University of Hamburg, D-22761 Hamburg, FRG
\item[\taiwan] National Central University, Chung-Li, Taiwan, China
\item[\tsinghua] Department of Physics, National Tsing Hua University,
      Taiwan, China
\item[\S]  Supported by the German Bundesministerium 
        f\"ur Bildung, Wissenschaft, Forschung und Technologie
\item[\ddag] Supported by the Hungarian OTKA fund under contract
numbers T019181, F023259 and T024011.
\item[\P] Also supported by the Hungarian OTKA fund under contract
  numbers T22238 and T026178.
\item[$\flat$] Supported also by the Comisi\'on Interministerial de Ciencia y 
        Tecnolog{\'\i}a.
\item[$\sharp$] Also supported by CONICET and Universidad Nacional de La Plata,
        CC 67, 1900 La Plata, Argentina.
\item[$\diamondsuit$] Also supported by Panjab University, Chandigarh-160014, 
        India.
\item[$\triangle$] Supported by the National Natural Science
  Foundation of China.
\item[\dag] Deceased.
\end{list}
}
\vfill


\newpage

%
%

\newpage

\end{document}